# Anomalous atmospheric circulation favored the spread of COVID-19 in Europe

Arturo Sanchez-Lorenzo[1*], Javier Vaquero-Martínez[1], Josep Calbó[2], Martin Wild[3,] Ana Santurtún[4], Joan-A. Lopez-Bustins[5], Jose-M. Vaquero[1], Doris Folini[3], Manuel Antón[1]

[1]*Department of Physics, University of Extremadura, Badajoz, Spain*

[2]*Department of Physics, University of Girona, Girona, Spain*

[3]*Institute for Atmosphere and Climate (IAC), ETH Zurich, Zurich, Switzerland*

[4]*Unit of Legal Medicine, Department of Physiology and Pharmacology, University of Cantabria, Santander, Spain*

[5]*Climatology Group, Department of Geography, University of Barcelona, Barcelona, Spain*

**Abstract**: The current pandemic caused by the coronavirus SARS-CoV-2 is having negative health, social and economic consequences worldwide. In Europe, the pandemic started to develop strongly at the end of February and beginning of March 2020. It has subsequently spread over the continent, with special virulence in northern Italy and inland Spain. In this study we show that an unusual persistent anticyclonic situation prevailing in southwestern Europe during February 2020 (i.e. anomalously strong positive phase of the North Atlantic and Arctic Oscillations) could have resulted in favorable conditions, in terms of air temperature and humidity, in Italy and Spain for a quicker spread of the virus compared with the rest of the European countries. It seems plausible that the strong atmospheric stability and associated dry conditions that dominated in these regions may have favored the virus's propagation, by short-range droplet transmission as well as likely by long-range aerosol (airborne) transmission.

**Background**

The world is currently undergoing a pandemic associated with the severe acute respiratory syndrome coronavirus 2 (SARS-CoV-2), which is a new coronavirus first noticed in late 2019 in the Hubei province, China[1,2]. The virus has a probable bat[3,4] origin, and causes the ongoing coronavirus disease 2019 (COVID-19). Although it is crucial to find a proper vaccine and medical treatment for this pandemic, it is also relevant to know the main factors controlling the transmission of the virus and disease, including the role of meteorological conditions in the spread of the virus. The World Health Organization (WHO) states that robust studies are needed to refine forecasting models and inform public health measures[5].

Respiratory virus infections can be transmitted via direct and indirect contact, or by means of particles (droplets or aerosols) emitted after a cough or sneeze or during conversation by an infected person. The large particles (>5 μm diameter) are referred to as respiratory droplets and tend to settle down quickly on the ground, usually within one meter of distance. The small particles (<5 μm in diameter) are referred to as droplet nuclei and are related to an airborne transmission. These particles can remain suspended in the air for longer periods of time and can reach a longer distance from the origin[6]. These small aerosol particles are inhalable and can penetrate all the way down to the alveolar space in the lungs[7], where cell receptors for some infectious respiratory viruses are located, including the angiotensin converting enzyme II (ACE2) used by SARS-CoV-2 to infect the individual[3].

Airborne transmission has been suggested to play a key role in some diseases like tuberculosis or measles, and even in coronaviruses[8–10]. A recent study has described that the SARS-CoV-2 virus can remain viable at least up to 3 hours in airborne conditions[11]. Respiratory droplets and aerosols loaded with pathogens can reach distances up to 7-8 meters under some specific conditions such as a turbulence gas cloud emitted after a cough of an infected person[12]. A study performed in Wuhan, the capital of the Hubei province, has shown that the SARS-CoV-2 virus could be found in several health care institutions, as well as in some crowded public areas of the city. It also highlights a potential resuspension of the infectious aerosols from the floors or other hard surfaces with the walking and movement of people[13]. Another study has also shown evidence of potential airborne transmission in a health care institution[14].

Recent studies have pointed out a main role of temperature and humidity in the spread of COVID-19. Warm conditions and wet atmospheres tend to reduce the transmission of the disease[15–22]. For example, it has also been pointed out that the main first outbreaks worldwide occurred during periods with temperatures around 5-11ºC, never falling below 0ºC, and specific humidity of 3-6 g/kg aproximately[18].

The first major outbreak in Europe was reported in Northern Italy in late February 2020. Following that, several major cases have been reported in Spain, Switzerland and France in early March, with a subsequent spread over many parts of Europe. At present (28th March 2020) Italy and Spain are still the two main contributors of cases and deaths in the continent, with major health, political and socio-economic implications. The main hypothesis of this work is that the atmospheric circulation pattern in February 2020 has helped to shape the spatial pattern of the outbreak of the disease in Europe.

**February 2020: a highly unusual circulation pattern**

The main atmospheric circulation pattern during February 2020 was characterized by an anomalous anticyclonic system over the western Mediterranean basin, centered between Spain and Italy, and lower pressures over Northern Europe centered over the Northern Sea and Iceland (Figure 1, Figure S1). This spatial configuration represents the well-known North Atlantic Oscillation (NAO)[23,24] in its positive phase, which is the teleconnection pattern linked to dry conditions in southern Europe whereas the opposite occurs in northern Europe[25].

Figure 2 and Figure S2 show maps for February 2020 for several meteorological fields that provide clear evidence of the stable atmospheric circulation in southern Europe, with a tendency towards very dry (i.e., lack of precipitation) and calm conditions. As suggested in an earlier analysis[18], the SARS-CoV-2 virus seems to be transmitted most effectively in dry conditions with daily mean air temperatures between around 5ºC and 11ºC, which are the conditions shown in Figure 2 for the major part of Italy and Spain. By contrast, northern Europe has experienced mainly wet and windy conditions due to an anomalous strong westerly circulation that is linked to rainy conditions. These spatial patterns fit with the well-known climate features associated over Europe during positive phases of the NAO[26]. The Arctic Oscillation (AO), which is a teleconnection pattern very much linked to NAO, showed in February 2020 the strongest positive value during 1950-2010 (Figure S3). The AO reflects the northern polar vortex variability at surface level[27], and it consists of a low-pressure centre located over the Norwegian sea and the Arctic ocean and a high-pressure belt between 40 and 50ºN, forming an annular-like structure. Positive values of the AO index mean a strong polar vortex, and the anomalous positive phase experienced during early 2020 has been linked with the recent ozone loss just registered over the Arctic region[28].

**Meteorology and the spatial pattern of the outbreak in Europe**

We argue that this spatial configuration of the atmospheric circulation might have played a key role in the modulation of the early spread of the COVID-19 outbreaks over Europe. It is known that some cases were reported already in mid-January in France, with subsequent cases in Germany and other countries[29]. Thus, the SARS-CoV-2 virus was already in Europe in early 2020, but it may only have started to extend rapidly when suitable atmospheric conditions for its spread were reached. It is possible that these proper conditions were met in February, mainly in Italy and Spain, due to the anticyclonic conditions previously mentioned.

The link between the COVID-19 spread and atmospheric circulation has been tested as follows. We have extracted the monthly anomalies of sea level pressure (SLP) and 500 hPa geopotential height for February 2020 over each grid point of the 15 capitals of the European countries (Figure S4) with the highest number of COVID-19 cases reported so far (see Data and Methods). Figure 3 (top) shows that there is a statistically significant

($R^2$=0.481, $p$<0.05) second order polynomial fit between the anomalies of the 500 hPa and the total cases per population. Italy, Spain, and Switzerland, which are the only countries with more than 1,000 cases/million inhabitants in our dataset, clustered together in regions with very large positive anomalies of 500 hPa geopotential heights. For the total number of deaths the fit is also statistically significant for a second order polynomial regression ($R^2$=0.50, $p$<0.05), and it shows clearly how Italy and Spain are out of scale compared to the rest of the European countries. Similar results are obtained using SLP fields (not shown).

These results evidence that it seems plausible that the positive phase of the NAO, and the atmospheric conditions associated with it, provided optimal conditions for the spread of the COVID-19 in southern countries like Spain and Italy, where both the start and the most severe impacts of the outbreak in Europe were located. To test this hypothesis further we have also analyzed the COVID-19 and meteorological data within Spain (see Data and Methods, Figure S5). The results show that mean temperature and specific humidity variables have the strong relation with COVID and fit with an exponential function (Figure 4). They indicate that lower mean temperatures (i.e., average of around 8-11°C) and lower specific humidity (e.g., <6 g/Kg) conditions are related to a higher number of cases and deaths in Spain. Nevertheless, it is worth mentioning that both meteorological variables are highly correlated ($R^2$=0.838, $p$<0.05) and are not independent of each other. The temperatures as low as 8-10°C are only reached in a few regions such as Madrid, Navarra, La Rioja, Aragon, Castille and Leon and Castilla-La Mancha. These areas are mainly located in inland Spain where drier conditions were reported the weeks before the outbreak. The rest of Spain experienced higher temperatures and consequently were out of the areas of higher potential for the spread of the virus, as reported so far in the literature[15–22]. In addition, higher levels of humidity also limit the impact of the disease, and therefore the coastal areas seem to benefit from lower rates of infection. Thus, in the southern regions of Spain (all of them with more than 13ºC and higher levels of specific humidity) we found lower rates of infection and deceases. This is in line with the spatial pattern in Italy, with the most (least) affected regions by COVID-19 mainly located in the North (South). In contrast, when the whole of Europe is considered on a country by country basis (see above and Figure 3), we find the opposite, a clear gradient with more severity from North to South as commented previously.

The spatial pattern of COVID-19 described above has some intriguing resemblances with the 1918 influenza pandemic, which is the latest deadly pandemic in modern history of Europe. The excess-mortality rates across Europe in the 1918 flu also showed a clear north-south gradient, with a higher mortality in southern European countries (i.e., Portugal, Spain or Italy) as compared to northern regions, an aspect that is not explained by socio-economic or health factors[30]. In Spain, a south-north gradient is also reported in the 1918 flu after controlling for demographic factors[31]. The central and northern regions of Spain experienced higher rates of mortality, and this has been suggested to be linked to more favorable climate conditions for influenza transmission as compared to the southern regions[31]. Interestingly, the SLP anomalies of the months before the major wave of this pandemic (which occurred in October-November 1918) shows a clear south-north dipole with positive anomalies in southern Europe centered over the Mediterranean, and negative ones in northern Europe (Figure S6). In other words, the NAO was also in its positive phase just before the major outbreak of the 1918 influenza pandemic. This resembles the spatial patterns described above for the current COVID-19 outbreak, both in terms of the spatial distribution of the mortality of the pandemic over Europe as well as in prevailing atmospheric circulation conditions before the major outbreak. These intriguing coincidences need further research in order to better understand the spatial and temporal distribution of large respiratory-origin pandemics over Europe.

Taking into account these results, we claim that the major initial outbreaks of COVID-19 in Europe (i.e., Italy and Spain) may be favored by an anomalous atmospheric circulation pattern in February, characterized by a positive phase of the NAO and AO. Taking into consideration current evidences in the literature, it seems that suitable conditions of air temperature and humidity were reached in Northern Italy and inland Spain. Indeed, meteorological conditions can affect the susceptibility of an infected host by altering the mucosal antiviral defense[32] and the stability and transmission of the virus[33], as well as social contact patterns[34]. We also hypothesize that the anomalous meteorological conditions experienced in Italy and Spain promoted the airborne contagion both indoors and outdoors, in addition to the direct and indirect contact and short-range droplets, which helped to speed up the rates of effective reproductive number (R) of the virus (Figure S7). Equally, the anticyclonic conditions, amplified in some areas by temperature inversions, may have reduced the dispersion of the virus outdoors. This stability and lack of precipitation can also produce more processes of suspension and resuspension of the infected aerosols indoors and, especially, outdoors, in a similar way as resuspension of anthropogenic pollutants in cities[35,36]. Equally, it is also suggested that high atmospheric pollutant concentrations can be positively related to increase fatalities related to respiratory virus infections[37,38] and even COVID-19[39]. This is a relevant issue as the main hotspot of COVID-19 in Italy is located in the Po valley, one of the most polluted regions of Europe, as well as the Madrid region (the most affected region so far in Spain)[40].

**Conclusions and outlook**

Although the outbreak of a pandemic is controlled by a high number of biological, health, political, social, economic and environmental factors, with complex and non-linear interrelationships between them, the results of this study indicate that an anomalous atmospheric circulation may explain why the COVID-19 outbreak in Europe developed more easily (or faster) in the south-west (mainly north of Italy and inland of Spain).

Specifically, the extreme positive phase of the AO and NAO during February 2020 could have modulated the beginning of the major outbreaks of COVID-19 in Europe. This detected anomalous atmospheric pattern, which produces dry conditions over southwestern Europe, may have provided optimal meteorological conditions for the virus propagation. In the context of anthropogenic climate change, it has been shown that in future emissions scenarios a poleward expansion of the Hadley cell is expected[41], which in turn is in line with a tendency to increase the frequency of positive phases of the NAO[42] (Figure S8). This should be taken into account for planning against future epidemics and pandemics that arise from respiratory viruses.

Interestingly, the conditions during the last major pandemic experienced in Europe (the Spanish flu in 1918), seem to resemble the current spatial pattern of affectation with more cases in the South of Europe as compared to the North. Equally, the dominant atmospheric situation was strongly affected by anticyclonic (cyclonic) conditions in the South (North) of

Europe. Further research is needed in order to better understand the spatio-temporal patterns of large epidemic and pandemic situations, and their connection with the prevailing atmospheric conditions patterns.

**References**

1. Huang, C. *et al.* Clinical features of patients infected with 2019 novel coronavirus in Wuhan , China. *Lancet* 497–506 (2020). doi:10.1016/S0140-6736(20)30183-5

2. WHO. *Novel coronavirus – China. Jan 12, 2020*. (2020).

3. Zhou, P. *et al.* A pneumonia outbreak associated with a new coronavirus of probable bat origin. *Nature* **579**, 270–273 (2020).

4. Liao, Y. *et al.* Identifying SARS-CoV-2 related coronaviruses in Malayan pangolins. https://doi.org/10.1038/s41586-020-2169-0 (2020).

5. WHO. *Coronavirus disease 2019 (COVID-19) Situation Report – 59*. (2020).

6. Gralton, J., Tovey, E., Mclaws, M. & Rawlinson, W. D. The role of particle size in aerosolised pathogen transmission : A review. *J. Infect.* **62**, 1–13 (2011).

7. CDC. *Coronavirus disease 2019 (COVID-19): how COVID-19 spreads. https://www.cdc.gov/coronavirus/2019-ncov/prevent-getting-sick/how-covid-spreads.html*. (2020).

8. Kutter, J. S., Spronken, M. I., Fraaij, P. L., Fouchier, R. A. M. & Herfst, S. Transmission routes of respiratory viruses among humans. *Curr. Opin. Virol.* **28**, 142–151 (2018).

9. Tellier, R., Li, Y., Cowling, B. J. & Tang, J. W. Recognition of aerosol transmission of infectious agents : a commentary. *BMC Infect. Dis.* 1–9 (2019).

10. Yu, I. T. S. *et al.* Evidence of Airborne Transmission of the Severe Acute Respiratory Syndrome Virus. *N. Engl. J. Med.* **350**, 1731–1739 (2004).

11. van Doremalen, N. *et al.* Aerosol and Surface Stability of SARS-CoV-2 as Compared with SARS-CoV-1. *N. Engl. J. Med.* (2020). doi:10.1056/NEJMc2004973

12. Bourouiba, L. Turbulent Gas Clouds and Respiratory Pathogen Emissions: Potential Implications for Reducing Transmission of COVID-19. *JAMA* (2020). doi:10.1001/jama.2020.4756

13. Liu, Y. *et al.* Aerodynamic Characteristics and RNA Concentration of SARS-CoV-2 Aerosol in Wuhan Hospitals during COVID-19 Outbreak. *bioRxiv* (2020). doi:10.1101/2020.03.08.982637

14. Santarpia, J. L. *et al.* Transmission Potential of SARS-CoV-2 in Viral Shedding Observed at the University of Nebraska Medical Center. *medRxiv* (2020). doi:10.1101/2020.03.23.20039446

15. Luo, W. *et al.* The role of absolute humidity on transmission rates of the COVID-19 outbreak. *medRxiv* under Rev. (2020). doi:10.1101/2020.02.12.20022467

16. B, O., L, C., Ferreira, N. & F, C. Role of temperature and humidity in the modulation of the doubling time of COVID-19 cases. *under Rev.* https://doi.org/10.1101/2020.03.05.20031872 (2020).

17. Wang, J., Tang, K., Feng, K. & Lv, W. High Temperature and High Humidity

Reduce the Transmission of COVID-19. *SSRN Electron. J.* 1–19 (2020). doi:10.2139/ssrn.3551767

18. Sajadi, M. M. *et al.* Temperature, Humidity and Latitude Analysis to Predict Potential Spread and Seasonality for COVID-19. *under Rev.* https://ssrn.com/abstract=3550308 (2020).

19. Bu, J. *et al.* Analysis of meteorological conditions and prediction of epidemic trend of 2019-nCoV infection in 2020. *medRxiv* (2020). doi:10.1101/2020.02.13.20022715

20. Chen, B. *et al.* Roles of meteorological conditions in COVID-19 transmission on a worldwide scale. *medRxiv* (2020). doi:10.1101/2020.03.16.20037168

21. Araujo, M. B. & Naimi, B. Spread of SARS-CoV-2 Coronavirus likely to be constrained by climate. *medRxiv* (2020). doi:10.1101/2020.03.12.20034728

22. Notari, A. Temperature dependence of COVID-19 transmission. *medRxiv* (2020). doi:10.1101/2020.03.26.20044529

23. Jones, P. D., Jonsson, T. & Wheeler, D. Extension to the North Atlantic oscillation using early instrumental pressure observations from Gibraltar and south- west Iceland. *Int. J. Climatol.* **17**, 1433–145 (1997).

24. Hurrell, J. W. Decadal Trends in the North Atlantic Oscillation: Regional Temperatures and Precipitation. *Science (80-. ).* **269**, 676–679 (1995).

25. Calbó, J. & Sanchez-Lorenzo, A. Cloudiness climatology in the Iberian Peninsula from three global gridded datasets (ISCCP, CRU TS 2.1, ERA-40). *Theor. Appl. Climatol.* **96**, 105–115 (2009).

26. Hurrell, J. W., Kushnir, Y., Ottersen, G. & Visbeck, M. An Overview of the North Atlantic Oscillation. in *The North Atlantic Oscillation: Climatic Significance and Environmental Impact* 1–35 (American Geophysical Union (AGU), 2003). doi:10.1029/134GM01

27. Baldwin, M. P. *et al.* Stratospheric Memory and Skill of Extended-Range Weather Forecasts. *Science* **301**, 636–640 (2003).

28. Witze, A. Rare ozone hole opens over Arctic — and it's big. *Nature* **580**, 18–19 (2020).

29. Spiteri, G. *et al.* First cases of coronavirus disease 2019 (COVID-19) in the WHO European Region, 24 January to 21 February 2020. *Eurosurveillance* **25**, (2020).

30. Ansart, S. *et al.* Mortality burden of the 1918-1919 influenza pandemic in Europe. *Influenza Other Respi. Viruses* **3**, 99–106 (2009).

31. Chowell, G., Erkoreka, A., Viboud, C. & Echeverri-Dávila, B. Spatial-temporal excess mortality patterns of the 1918-1919 influenza pandemic in Spain. *BMC Infect. Dis.* **14**, 1–12 (2014).

32. Kudo, E. *et al.* Low ambient humidity impairs barrier function and innate resistance against influenza infection. *Proc. Natl. Acad. Sci.* **116**, 10905–10910 (2019).

33. Moriyama, M. & Hugentobler, W. J. Seasonality of Respiratory Viral Infections. *Annu. Rev. ofVirology* **7**, 1–19 (2020).

34. Willem, L., van Kerckhove, K., Chao, D. L., Hens, N. & Beutels, P. A Nice Day for an Infection? Weather Conditions and Social Contact Patterns Relevant to Influenza

Transmission. *PLoS One* **7**, (2012).

35. Pant, P. & Harrison, R. M. Estimation of the contribution of road traffic emissions to particulate matter concentrations from field measurements: A review. *Atmos. Environ.* **77**, 78–97 (2013).

36. Querol, X. *et al.* Speciation and origin of PM10 and PM2.5 in Spain. *J. Aerosol Sci.* **35**, 1151–1172 (2004).

37. Cui, Y. *et al.* Environmental Health : A Global Air pollution and case fatality of SARS in the People ' s Republic of China : an ecologic study. **5**, 1–5 (2003).

38. Chen, G. *et al.* The impact of ambient fi ne particles on in fl uenza transmission and the modi fi cation effects of temperature in China : A multi-city study. *Environ. Int.* **98**, 82–88 (2017).

39. Wu, X., Nethery, R. C., Sabath, B. M., Braun, D. & Dominici, F. Exposure to air pollution and COVID-19 mortality in the United States. *medRxiv* (2020). doi:10.1101/2020.04.05.20054502

40. EEA. *Air quality in Europe — 2019 report*. (2019). doi:doi:10.2800/822355

41. Collins, M. *et al.* Long-term Climate Change: Projections, Commitments and Irreversibility. in *Climate Change 2013: The Physical Science Basis. Contribution of Working Group I to the Fifth Assess- ment Report of the Intergovernmental Panel on Climate Change* (ed. Stocker, T.F., D. Qin, G.-K. Plattner, M. Tignor, S.K. Allen, J. Boschung, A. Nauels, Y. Xia, V. B. and P. M. M.) 1029–1136 (Cambridge University Press, Cambridge, United Kingdom and New York, NY, USA, 2013).

42. Deser, C., Hurrell, J. W. & Phillips, A. S. The role of the North Atlantic Oscillation in European climate projections. *Clim. Dyn.* **49**, 3141–3157 (2017).

**Data and methods**

NCEP/NCAR[1], ERA5[2] and ERA20C[3] atmospheric data are used in this manuscript. The maps and data have been retrieved by using the tools and websites referenced in the main text, and more details about the spatial and temporal resolution, vertical levels, assimilation schemes, etc. can be consulted in their references. In brief, an atmospheric reanalysis like those used here is a climate data assimilation project which aims to assimilate historical atmospheric observational data spanning an extended period, using a single consistent assimilation scheme throughout, with the aim of providing continuous gridded data for the whole globe.

The Artic Oscillation (AO) index has been extracted from the Climate Prediction Center of the National Oceanic and Atmospheric Administration (NOAA). The AO index is constructed by projecting the daily 1000 hPa height anomalies poleward of 20°N onto the loading pattern of the AO, this latter being defined as the leading mode of Empirical Orthogonal Function (EOF) analysis of monthly mean 1000 hPa height. More details and data: https://www.cpc.ncep.noaa.gov/products/precip/CWlink/daily_ao_index/ao.shtml.

COVID-19 data on country basis were obtained on March 26$^{th}$, 2020 from the website https://www.worldometers.info/coronavirus/, which it is mainly based on the data provided by the Coronavirus COVID-19 Global Cases by the Center for Systems Science and Engineering (CSSE) at the Johns Hopkins University. Data from Spain on regional scale were obtained on March 28th, 2020 from the Spanish Government through the Institute of Health Carlos III (ISCIII): https://covid19.isciii.es/

For the link between the COVID-19 spread on European scale and atmospheric circulation we have extracted the monthly anomalies of sea level pressure (SLP) and 500 hPa geopotential height for February 2020 over each grid point of the 15 capitals of the European countries. We have selected the SLP and 500 hPa fields in order to summarize the meteorological conditions over each location, as it is known that several meteorological variables can be involved in the transmission of respiratory viruses[4,5]. With this approach we also avoid the lack of properly updated data for all potential meteorological variables involved in the COVID-19 spread, which needs further research as soon as the pandemic ends and a more reliable and complete database of both COVID-19 impact and meteorological data can be compiled[6].

For Spain, several meteorological variables with high-quality records were obtained from the Spanish Meteorology Agency (AEMET) based on surface observations for each of the capital cities of the provinces inside each autonomous region. Specifically, monthly averages for February 2020 of 2-m temperature, 2-m maximum temperature, 2-m minimum temperature (°C), air pressure (hPa), wind speed (km h$^{-1}$), specific humidity (g Kg$^{-1}$), relative humidity (%), total precipitation (mm) and days of more than 1 mm of precipitation. An arithmetic average has been calculated for the autonomous regions with more than one province.

We have checked Coupled Model Intercomparison Project Phase 5 (CMIP5) simulations for two future scenarios (RCP4.5 and RCP8.5) at the end of the 21th century.

Statistical analysis was performed with R software package for linear, multiple and polynomial regressions. The statistical significance was considered at the 5% level of confidence ($p<0.05$).

1.  Kalnay, E. *et al.* The NCEP / NCAR 40-Year Reanalysis Project. *Bull. Am. Meteorol. Soc.* **77**, 437–470 (1996).

2.  Copernicus Climate Change Service (C3S). *ERA5: Fifth generation of ECMWF atmospheric reanalyses of the global climate. Copernicus Climate Change Service Climate Data Store (CDS). https://cds.climate.copernicus.eu/cdsapp#!/home*. (2017).

3.  Poli, P., Hersbach, H., Berrisford, P. & other authors. ERA-20C Deterministic. https://www.ecmwf.int/node/11700. *ERA Report, 20* 48 (2015).

4.  Lowen, A. C., Mubareka, S., Steel, J. & Palese, P. Influenza Virus Transmission Is Dependent on Relative Humidity and Temperature. *PLOS Pathog.* **3**, 1–7 (2007).

5.  Fuhrmann, C. The effects of weather and climate on the seasonality of influenza: What we know and what we need to know. *Geogr. Compass* **4**, 718–730 (2010).

6.  Araujo, M. B. & Naimi, B. Spread of SARS-CoV-2 Coronavirus likely to be constrained by climate. *medRxiv* (2020). doi:10.1101/2020.03.12.20034728

**Acknowledgments**

A. Sanchez-Lorenzo was supported by a fellowship RYC-2016–20784 funded by the Ministry of Science and Innovation. Javier Vaquero-Martinez was supported by a predoctoral fellowship (PD18029) from Junta de Extremadura and European Social Fund. J.A. Lopez-Bustins was supported by Climatology Group of the University of Barcelona (2017 SGR 1362, Catalan Government) and the CLICES project (CGL2017-83866-C3-2-R, AEI/FEDER, UE). This research was supported by the Economy and Infrastructure Counselling of the Junta of Extremadura through grant GR18097 (co-financed by the European Regional Development Fund). J. Vázquez, Xavi B. and Raúl J.I. (UPV/EHU) kindly helped us in discussing the results. NCEP Reanalysis data provided by the NOAA/OAR/ESRL PSL, Boulder, Colorado, USA, from their Web site at https://psl.noaa.gov/

**Author's contribution**

A.S.L., J.V.M., J-M.V. and M.A. designed the research. A.S.L., J.V.M and J-A.L.B conducted the analyses. A.S.L., J.C., M.W., A.S. and M.A. refined the interpretations. A.S.L. wrote the manuscript. J.V.M, J.C., M.W., A.S., J-A.L.B, J-M.V. and M.A. provided comments and contributed to the text.

**Supplementary information** accompanies this paper

**Competing financial interests**: The authors declare no competing financial interests

**Corresponding author**: Arturo Sanchez Lorenzo, Department of Physics University of Extremadura, Badajoz, Spain, E-mail: arturosl@unex.es

# Figures

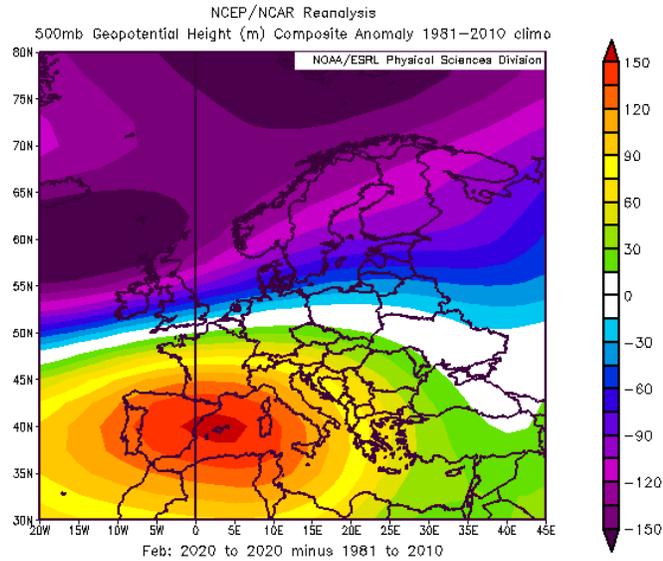

Figure 1. Anomaly pattern of 500 hPa geopotential height (m) for February 2020 over Europe as compared to the climatology mean (1981-2010 period). Image generated with the Web-based Reanalysis Intercomparison Tool provided by the NOAA/ESRL Physical Sciences Laboratory, Boulder Colorado from their Web site at http://psl.noaa.gov/

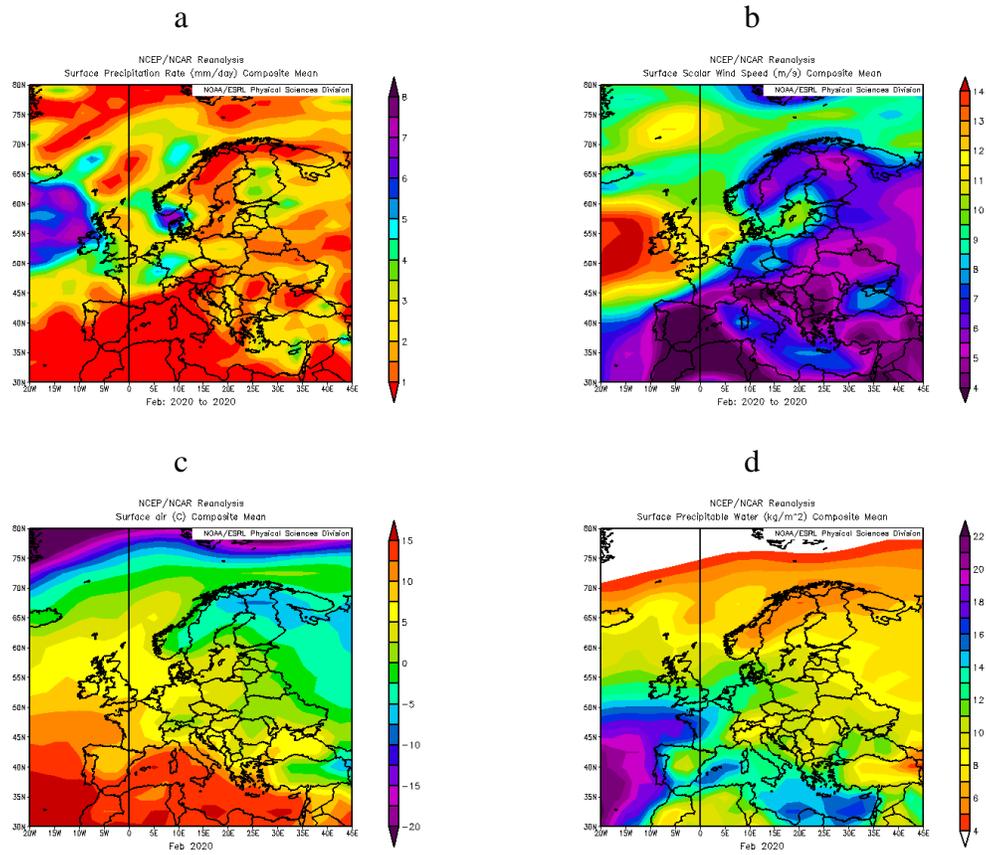

Figure 2. Mean values of several meteorological variables for February 2020 over Europe. a) Precipitation rate (mm/day), b) Surface wind speed (m/s), c) Surface air temperature (ºC), and d) Precipitable water (kg/m$^2$). Image generated with the Web-based Reanalysis Intercomparison Tool provided by the NOAA/ESRL Physical Sciences Laboratory, Boulder Colorado from their Web site at http://psl.noaa.gov/

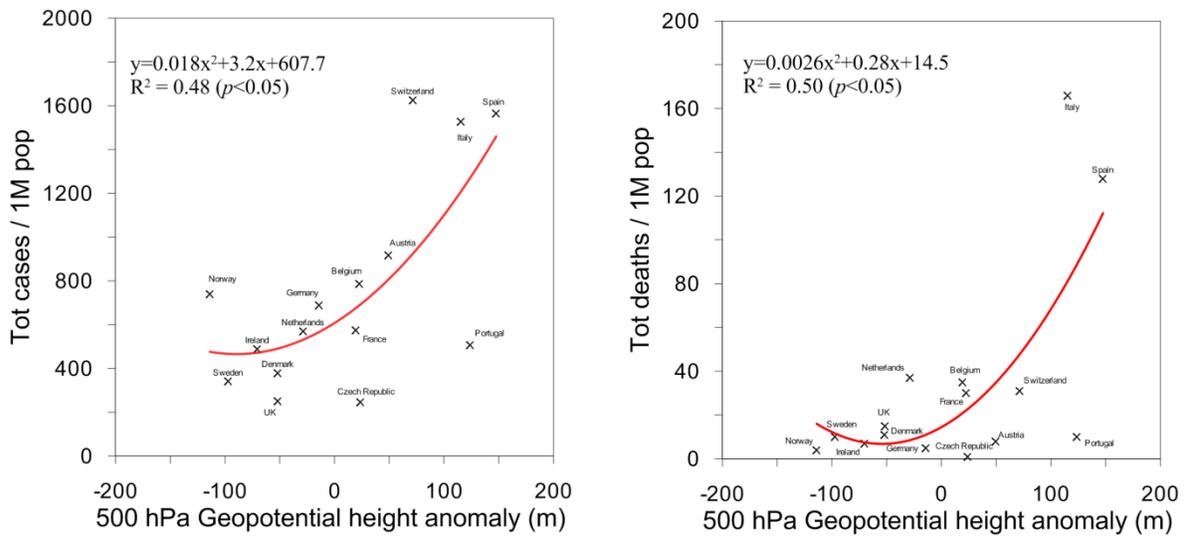

Figure. 3. Relationship between COVID-19 cases and deaths in Europe and 500 hPa geopotential height anomalies (m) over the capital of each country. Each point represents one of the 15 countries with more cases reported up to March 26[th], 2020. The 500 hPa geopotential height anomalies are calculated for February 2020 with respect to the 1981-2010 climatological mean.

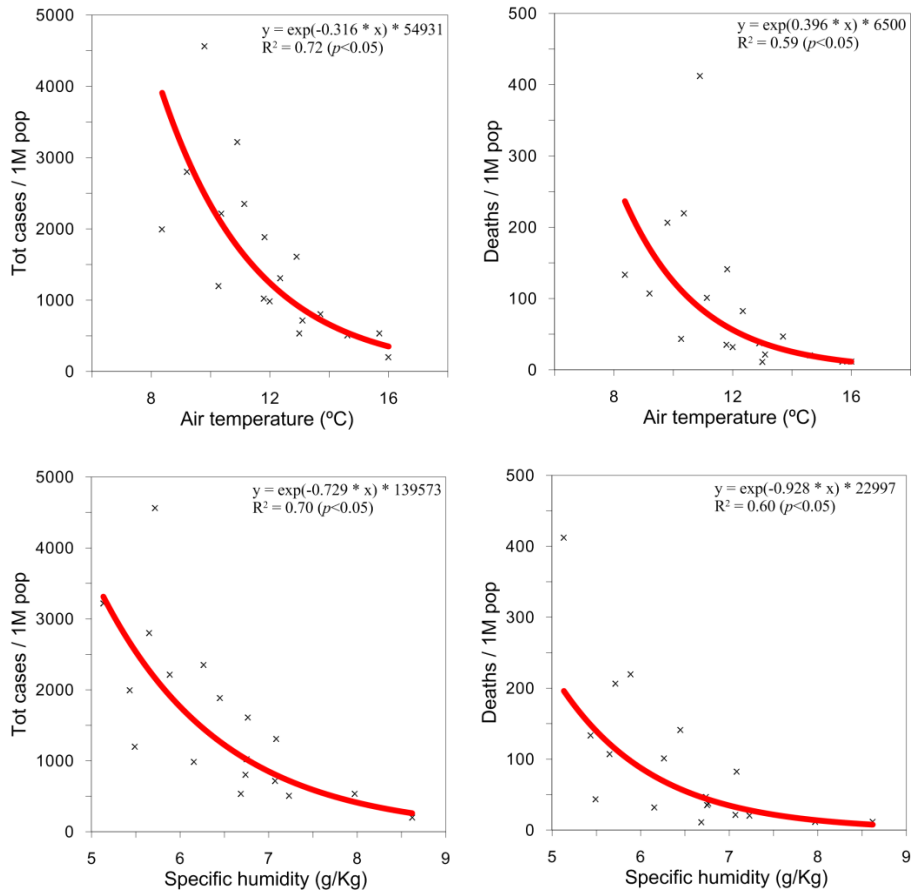

Figure 4. Relationship between mean (top) air temperature (ºC) and (bottom) specific humidity (g/Kg) against COVID-19 cases (left) and deaths (right) in Spain as reported up to March 28[th], 2020. Each cross indicates a region of Spain. The meteorological data refer to the average of February 2020.

# Supplementary Information

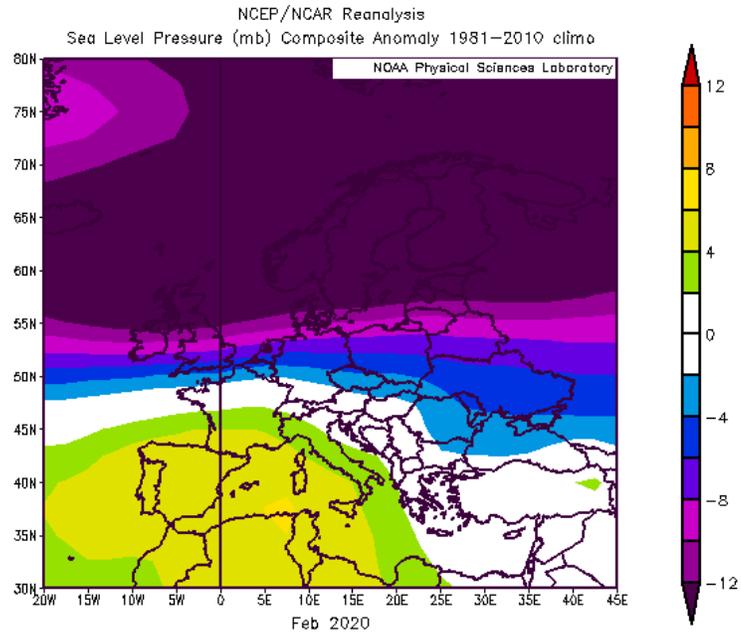

Figure S1. Anomaly pattern of sea level pressure (mb) for February 2020 over Europe as compared to the climatology mean (1981-2010 period). Image generated with the Web-based Reanalysis Intercomparison Tool provided by the NOAA/ESRL Physical Sciences Laboratory, Boulder Colorado from their Web site at http://psl.noaa.gov/

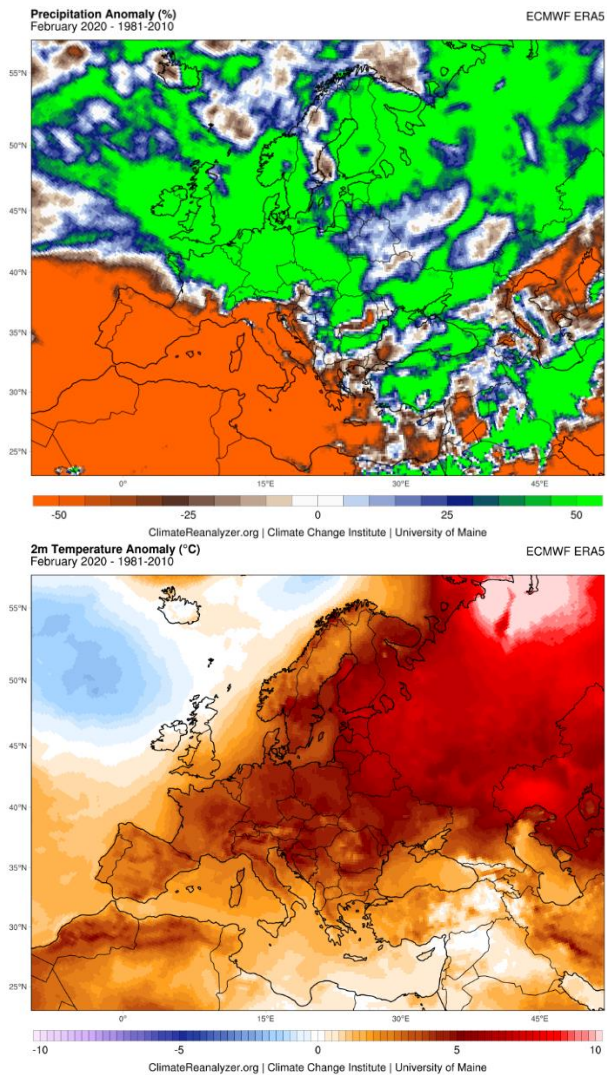

Figure S2. Anomaly pattern of (top) precipitation (in %) and (bottom) 2-m temperature (in ºC) during February 2020 as compared to the climatology mean (1981-2010 period). Image generated with Climate Reanalyzer, Climate Change Institute, University of Maine, USA (https://ClimateReanalyzer.org).

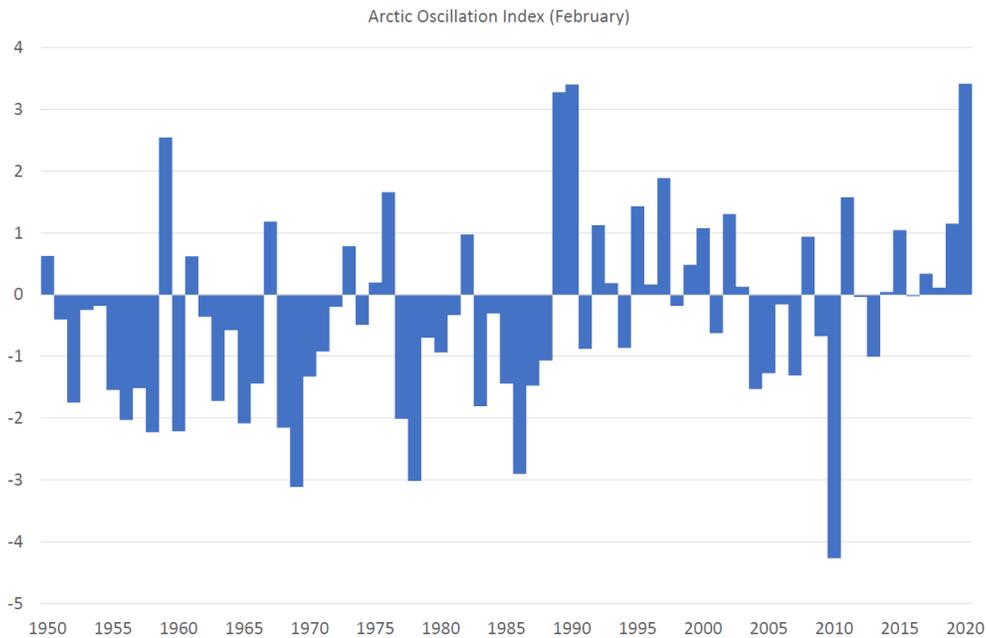

Figure S3. Time series of the Artic Oscillation (AO) index from 1950 to 2020 provided by the Climate Prediction Center of the National Oceanic and Atmospheric Administration (NOAA). The February 2020 is the highest value (3.417), closely followed by 1990 (3.402) and 1989 (3.279) and far away from the 4[th] largest value in 1959 (2.544). This highlights the exceptionally extreme AO pattern of the past February 2020 in a long-term perspective.

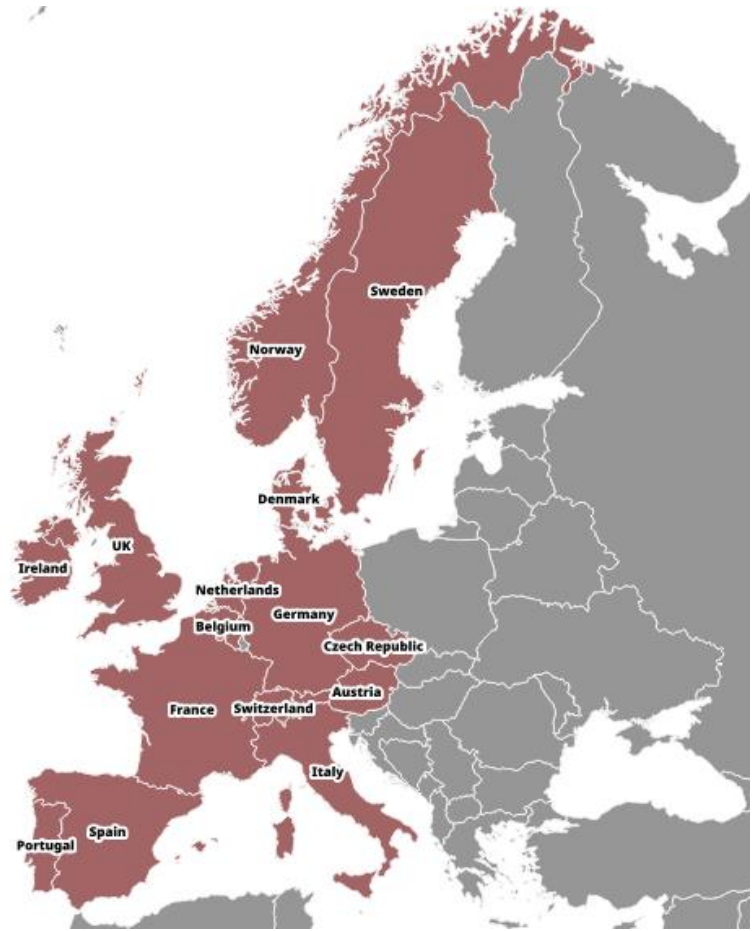

Figure S4. Location of the 15 countries used in this study that provided cases and deaths of COVID-19.

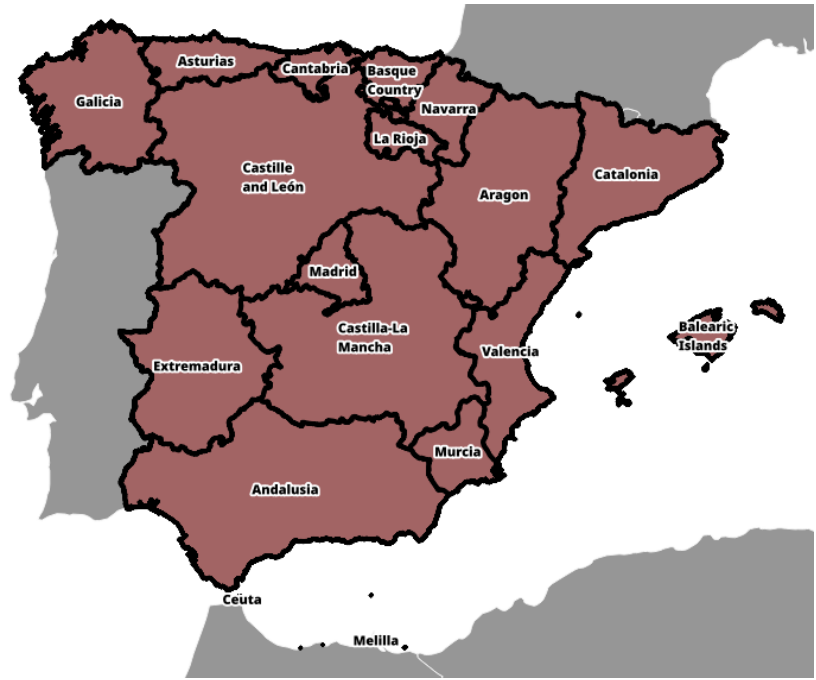

Figure S5. Location of the autonomous communities of Spain, as well as the two autonomous cities of Ceuta and Melilla. The Canary Islands has not been included in this study due to its geographical location in tropical latitudes.

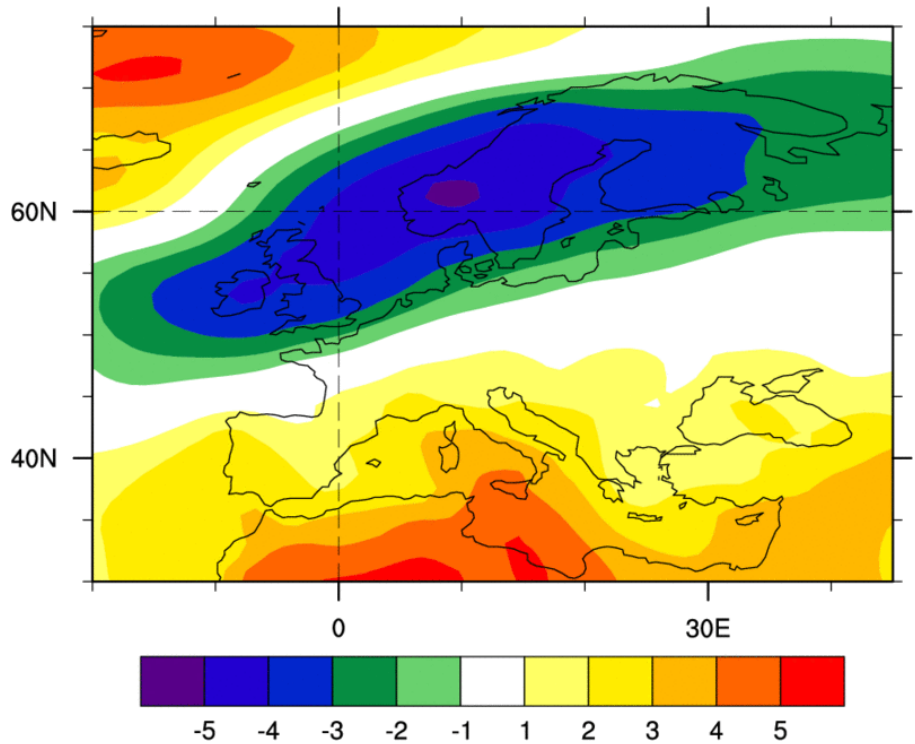

Figure S6. Anomaly map of the sea level pressure (SLP) field extracted from ERA20C reanalysis of September and October 1918 as compared to the climatological mean (1981-2010 period). Image generated with the Web-based Reanalysis Intercomparison Tool provided by the NOAA/ESRL Physical Sciences Laboratory, Boulder Colorado from their Web site at http://psl.noaa.gov/

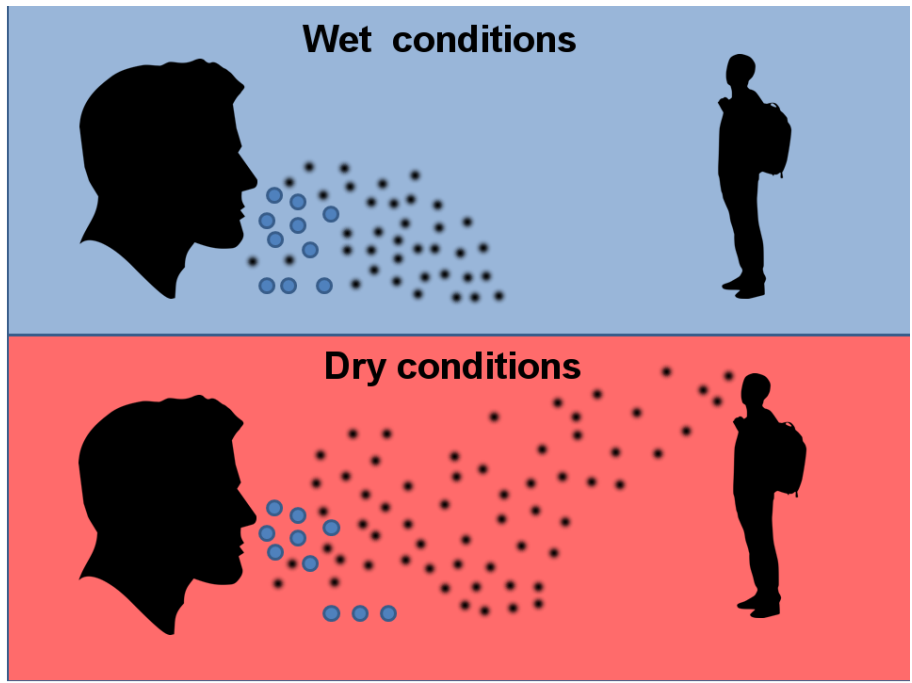

Figure S7. Schematic representation of particles emitted by a cough, with the large droplets settled down nearby (e.g., 1 m distance) and the smaller airborne particles spreading in suspension for longer time, and reaching longer distances, especially in dry and stable conditions as compared to wet environments. It is also possible that a resuspension of aerosol particles can eventually happen due to human activities (e.g., walking, cleaning, etc.) or air flows, which is enhanced under dry conditions due to the lack of precipitation.

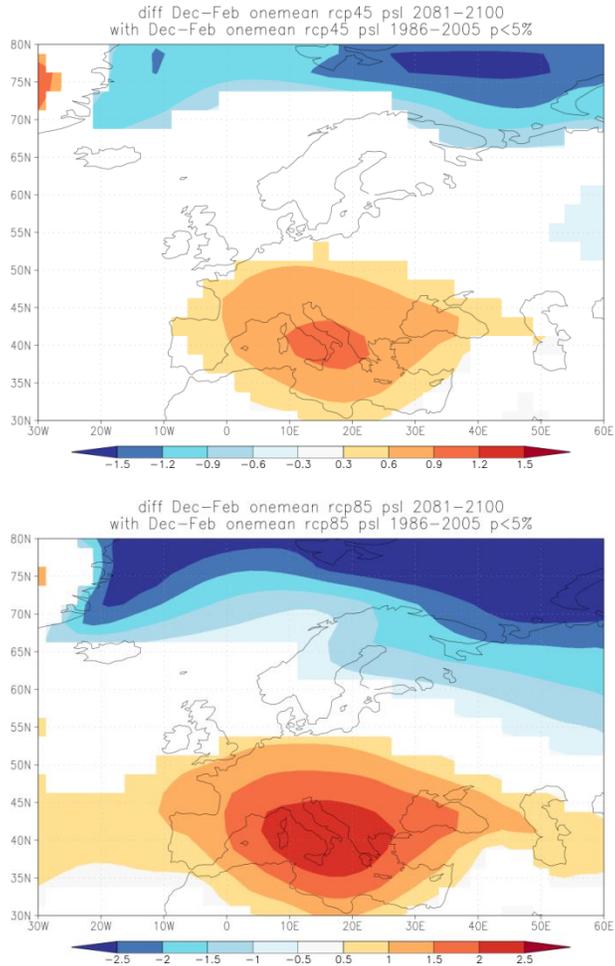

Figure S8. Winter (DJF) anomalies of sea level pressure (SLP, in mb) as computed for the CMIP5 ensemble (one member per model) for the (top) RCP4.5 and (bottom) RCP8.5 emission scenarios at the end of the 21st century. The differences are computed expressed as the 2081-2100 minus 1986-2005 period. Only statistically significant fields ($p<0.05$) are plotted as estimated by a Student's t-Test. Map composed with the data and tools provided by the KNMI Climate Explorer website (https://climexp.knmi.nl/start.cgi). The maps show a consistent picture of an intensification of the positive NAO phase, which implies that in the future winter conditions as experienced over Europe past February 2020 could become more common.